\documentclass[12pt]{article}
\newcommand{\be}{\begin{equation}}
\newcommand{\ee}{\end{equation}}
\newcommand{\bea}{\begin{eqnarray}}
\newcommand{\eea}{\end{eqnarray}}
\newcommand{\ba}{\begin{array}}

\newcommand{\ea}{\end{array}}

\def\bbox{{\,
\lower0.9pt\vbox{\hrule \hbox{\vrule height 0.2 cm
\hskip 0.2 cm \vrule height 0.2 cm}\hrule}\,}}
\newcommand{\dsl}{\pa \kern-0.5em /}

\newcommand{\nn}{\nonumber \\}

\def\ds{\raise.15ex\hbox{/}\kern-.57em\partial}
\def\Ds{\,\raise.15ex\hbox{/}\mkern-13.5mu D}
\begin{document}

\baselineskip 18pt


\begin{titlepage}
\vfill
\begin{flushright}
KIAS-P01052\\
hep-th/0111199\\
\end{flushright}

\vfill

\begin{center}
\baselineskip=16pt 
{\Large\bf D-Brane Probe and Closed String Tachyons} \\
\vskip 10mm
{Yoji Michishita$^\&$ and Piljin Yi$^\#$}
\vskip 0.5cm
{\small\it
School of Physics, Korea Institute for Advanced Study\\
207-43, Cheongryangri-Dong, Dongdaemun-Gu, Seoul 130-012, Korea}
\end{center}
\vfill
\par
\begin{center}
{\bf ABSTRACT}
\end{center}
\begin{quote} 
We consider a D-brane probe in unstable string background associated with 
flux branes. The twist in spacetime metric reponsible for the supersymmetry
breaking is shown to manifest itself in mixing of open Wilson lines with the 
phases of some adjoint matter fields, resulting in a non-local and
nonsupersymmetric form of
Yang-Mills theory as the probe dynamics. This provides a setup where one can
study fate of a large class of unstable closed string theories that include
as a limit type 0 theories and various orbifolds of type II and type 0 
theories. We discuss the limit of ${\bf C}/Z_n$ orbifold in some detail and 
speculate on couplings with closed string tachyons.
\end{quote}

\vfill
\vskip 5mm
\hrule width 5.cm
\vskip 5mm
\begin{quote}
{\small
\noindent $^\&$ E-mail: michishi@kias.re.kr\\
\noindent $^\#$ E-mail: piljin@kias.re.kr\\
}
\end{quote}
\end{titlepage}
\setcounter{equation}{0}

\section{Introduction and Summary}

Recently, 
study of unstable string theories \cite{dd,Yi,emil,osft,bsft,vacuum} has 
added important new understanding of string theory in general.
General philosophy is nothing new. One looks for tachyonic degrees of 
freedom in perturbative string sector and interprets its condensation as
a sign that the unstable theory is moving toward a more stable theory.
In some cases, such as open string sector associated with unstable D-branes,
there are fairly convincing arguments that the end point of the condensation
is a supersymmetric vacuum. These studies were possible in part because
open strings per se do not contain gravitational degrees of freedom. 
The background geometry of spacetime, for instance, may be regarded as
given, and one studies a field theory of some sort in that given background.

When it comes to closed string theories with tachyonic degrees of freedom,
things are more involved. Since closed strings contain gravity at tree level,
condensation of tachyon induces immediate reaction by the spacetime
geometry, study of which seems considerably more difficult. A recent work
by Adams et. al.\cite{adams} tried to alleviate this problem by considering 
unstable closed string theories, but with tachyons living in lower dimensional
subspace, namely at orbifold fixed points. Effect of tachyon condensation
on spacetime geometry is studied via D-brane probe analysis and also from
beta function analysis, which seems to point toward a supersymmetric
string theory in a flat background as the end point.
The models they considered are nonsupersymmetric and noncompact 
orbifolds of 2 and 4 dimensions. 

In this note, we would like to offer D-brane probe analysis of a continuous
family of unstable closed string theories, which include the orbifolds of
Adams et. al. as special limits \cite{suyama,takayanagi1}. 
Consider type II string theory on an orbifold of type
\be
 {\bf R}^{1+6} \times ({\bf R}^2\times {\bf R}^1)/Z ,
\ee
where the integer group acts as a rotation on ${\bf R}^2$ and a shift on 
${\bf R}^1$.
Denoting the coordinates on each factor as $re^{i\phi}=x^7+ix^8$ and 
$y=x^9$, we define the action as
\be
\phi\leftrightarrow\phi +2\pi \epsilon,\qquad y\leftrightarrow y+2\pi R .
\ee
The string spectrum in such background has been analyzed by Russo and
Tseytlin  \cite{russo2}, 
who found tachyonic degrees of freedom for sufficiently small
$R$. Type IIA string in this background is related to type 
IIA string
on the so-called Melvin universe via 9-11 flip, and is also related to 
type 0 theory  \cite{gutperle1,gutperle2,russo3}. 
For instance, $R\rightarrow 0$ limit with $\epsilon=1$
is supposed to yield a type 0 string theory in flat 10D background.

In order to keep an angular coordinate of period $2\pi$,
we may introduce a new coordinate by $\phi=\theta+b\, y$ with 
$b\equiv \epsilon/R$. With respect to the new coordinate system, 
the metric of the spacetime is
\be
-(dt)^2+(dx^1)^2+\cdots+(dx^6)^2+dr^2+r^2(d\theta+ b\, dy)^2 +(dy)^2 ,
\ee
which is locally flat.

According to the perturbative string spectrum of these models \cite{russo2}, 
tachyons reside in winding sector(s) along $y$.
The fact that tachyons are in winding sectors also implies that 
tachyons are confined to the region near $r=0$; 
A string that winds around $y$ must also wander in ${\bf R}^2$
and traverse a net angle of $2\pi\epsilon$. Because of this,
a string is longer and thus heavier when
its endpoints are located away from the origin. In effect,
the tachyon lives in a lower 
dimensional subspace, and one may hope that D-brane probe
analysis may yield some answers on the fate of this unstable theory upon
tachyon condensation. In this note, we construct and analyze D-brane
probe theory in the above background for general twist, and try to make
contact with various limits such as  type 0 theories and 
nonsupersymmetric orbifolds of type II theories.
For CFT construction of D-branes on this background, see 
 \cite{dudas,takayanagi2}.

For convenience, let us summarize the probe theory action. 
We probe the background
by $N$ D0 branes. The resulting worldvolume theory is a 2-dimensional
$U(N)$ Yang-Mills theory on the dual circle of circumference $1/R$.
There are eight adjoint Higgs fields denoted by six hermitian
${\cal X}^I$'s corresponding to ${\bf R}^6$ and a complex field and its
conjugate, $\cal Z$ and $\bar{\cal Z}$ associated with
${\bf R}^2$.
Fermions are in an 16 component spinor $\Psi$ also in the adjoint. 
The Lagrangian is formally identical to that of standard $N=8$,
maximally supersymmetric Yang-Mills theory 
\bea
\int dt d\sigma&&\hskip -5mm
\frac 12{\rm tr}\biggl(-{\cal F}_{01}^2-\sum_I([{\cal D},
{\cal X}^I])^2 -|[{\cal D},{\cal Z}]|^2 \nn
&&+\sum_{I>J}[{\cal X}^I,
{\cal X}^J]^2+\sum_I[{\cal X}^I,{\cal Z}][{\cal X}^I,\bar{\cal Z}]+
[{\cal Z},\bar {\cal Z}] [\bar{\cal Z},{\cal Z}]/4 \nn
&&+i\bar\Psi\Gamma^i[{\cal D}_i,\Psi]  -\sum_I\bar\Psi\Gamma_I 
[{\cal X}^I,\Psi]
-\bar\Psi\Gamma_{\bar Z} [\bar{\cal Z},\Psi]-
\bar\Psi\Gamma_{ Z} [{\cal Z},\Psi]\biggr),
\eea
with $SO(9,1)$ Dirac matrices $\Gamma_{0,1,2,...,9}$
except for one important detail: $\cal Z$ and $\Psi$ are always accompanied 
by a Wilson line. For instance, everywhere in the above expressions, 
one should replace $\cal Z$ by
\be
{\cal Z}(\sigma)\rightarrow 
[P e^{-i\int_{\sigma}^{\sigma-b}{\cal A}}\,]\,{\cal Z}(\sigma-b) ,
\ee
where $\cal A$ is the gauge field along the dual circle. Also,
breaking up the Fermions in terms of chirality with respect to 
${\bf R}^2$, and calling them $\Psi_\pm$, one must make
the replacement
\be
\Psi_\pm(\sigma)\rightarrow
[P e^{-i\int_{\sigma}^{\sigma\mp b/2}{\cal A}}\,]\,\Psi_\pm(\sigma\mp b/2) .
\ee
 Subsequently, the argument of the field that multiplies to the right (to 
the left  of $\bar{\cal Z}$ and of $\bar \Psi$) must be accordingly shifted 
as is necessary to  maintain covariance.

\section{D0 Probe}

\subsection{Generality}

Let us probe this background by D0 branes. Following Taylor \cite{taylor}, 
we may derive
the effective dynamics of D0's by introducing infinite number of images
upon the action of $Z$. Introducing Chan-Paton factor $|n\rangle$, the lowest
lying mode of open strings ending on the D0's can be written as infinite 
matrices. Denote the 9 transverse scalars by upper case symbols,
\bea
&& X^{I}_{mk},\quad I=1,2,3,4,5,6 ,\nn
&& Z_{mk},\nn
&& \bar Z_{mk},\nn
&& Y_{mk}.
\eea
The matrices satisfy a reality condition, which requires in particular,
\be
\bar Z_{mk}= Z_{km}^*,
\ee
while $Y$ and $X^I$ are Hermitian. 

Action of the orbifold group is such that\footnote{This action can also
be found in Ref.~\cite{motl}.}
\bea
Y_{m+1,k+1} &\leftrightarrow& Y_{mk} + 2\pi R \delta_{mk},\nn
Z_{m+1,k+1}&\leftrightarrow& e^{2\pi i\epsilon}Z_{mk},\nn
\bar Z_{m+1,k+1}&\leftrightarrow& e^{-2\pi i\epsilon}\bar Z_{mk},\nn
X^I_{m+1,k+1}&\leftrightarrow&X^I_{mk},
\eea
where one should take each component as representing a square block.
The block would be $N\times N$ if there are $N$ D-brane probes.
Configurations invariant under such mapping can be solved for and give
\bea
Y_{mk}&=&{\cal A}_{m-k}+2\pi m R\,\delta_{mk},\nn
Z_{mk}&=&f_{m-k}^{(\beta)} 
e^{\pi i(m-k)\epsilon\beta}e^{\pi i(m+k)\epsilon},\nn
X^I_{mk}&=&{\cal X}^I_{m-k},
\eea
where $\beta$ is an arbitrary real number. This number does not enter the
physics and we will later take it to be $-1$ as a matter of convenience.

Reality condition implies
that 
\bea
{\cal A}_m&=&({\cal A}_{-m})^*,\nn
{\cal X}^I_{m}&=&({\cal X}^I_{-m})^* ,
\eea
and also that
\bea
\bar Z_{mk}=(Z_{km})^\dagger=(f_{k-m}^{(\beta)})^\dagger
e^{\pi i(m-k)\epsilon\beta}e^{-\pi i(m+k)\epsilon}.
\eea
Finally, gauge transformation is also reduced similarly to satisfy,
\be
U_{m+1,k+1} = U_{mk}.
\ee
In principle, one could derive low energy effective action of D-branes
by restricting open string fields to the above constraints.

A little more illuminating procedure can be found by trading off the
discrete indices in favor of a continuous variable on a circle of 
circumference $1/R$. In more concrete terms, we define
\be
C(\sigma,\sigma')\equiv R\sum_{mk} C_{mk} e^{-2\pi i R m \sigma}
e^{2\pi i R k \sigma'},
\ee
for each object in adjoint representation, and find
\bea
Y(\sigma,\sigma')&=&[{\cal A}(\sigma)+i\partial_\sigma ]\,
\delta(\sigma-\sigma'),\nn
Z(\sigma,\sigma')&=&f(\sigma) 
\,\delta(\sigma-\sigma'-\epsilon/R),\nn
\bar Z(\sigma,\sigma')
&=&f(\sigma')^\dagger 
\,\delta(\sigma'-\sigma-\epsilon/R), \nn
X^I(\sigma,\sigma') &=&{\cal X}^I(\sigma)\,\delta(\sigma-\sigma'),
\eea
with $\beta=-1$ suppressed in the notation. The delta function is normalized by
\be
\int_0^{1/R} \delta(\sigma-\sigma_0)\, d\sigma=1 .
\ee
Similarly we find gauge transformation is encoded in a local quantity,
\be 
U(\sigma,\sigma')=U(\sigma)\,\delta(\sigma-\sigma').
\ee
With these, it is not difficult to convince oneself that all of the above 
fields transform under the gauge transformation bilocally. That is,
\be
C(\sigma,\sigma')\Rightarrow  
U(\sigma)\,C(\sigma,\sigma')\,U(\sigma')^\dagger ,
\ee
which is in reality local except for $f$. That is,
\bea
[{\cal A}(\sigma)+i\partial_\sigma ]&\Rightarrow&
U(\sigma)[{\cal A}(\sigma)+i\partial_\sigma ]U(\sigma)^\dagger, \nn
{\cal X}^I(\sigma)&\Rightarrow&U(\sigma){\cal X}^I(\sigma)U(\sigma)^\dagger ,
\eea
where
${\cal A}(\sigma)$ is the gauge field on the dual circle 
and transforms as such. On the other hand, we have
\bea
f(\sigma)&\Rightarrow& U(\sigma)f(\sigma) 
U(\sigma-\epsilon/R)^\dagger .
\eea
We can trade off this $f$ in favor of ${\cal Z}$ 
satisfying
\be
{\cal Z}(\sigma)\Rightarrow U(\sigma){\cal Z}(\sigma)U(\sigma)^\dagger  ,
\ee
with a field redefinition involving open Wilson line
\be
W(\sigma ,\sigma')\equiv P e^{-i\int_{\sigma}^{\sigma'}{\cal A}} ,
\ee
such that
\bea
Z(\sigma,\sigma')&=& W(\sigma,\sigma-\epsilon/R)\,{\cal Z} 
(\sigma-\epsilon/R)\,
\delta(\sigma-\sigma'-\epsilon/R) ,
\eea
and similarly
\bea
\bar Z(\sigma,\sigma') &=&{\cal Z}^\dagger (\sigma)\,
 W(\sigma,\sigma+\epsilon/R)\,
\delta(\sigma-\sigma'+\epsilon/R) .
\eea

\subsection{Commutators and the Probe Theory}

Matrix multiplications involving $Y$, $Z$, $X^I$ can be mapped to
integrals along the dual circle as follows,
\be
\sum_l C_{mk}D_{kl} \quad\rightarrow \quad
\int d\sigma'' C(\sigma,\sigma'')D(\sigma'',
\sigma') .
\ee
An extra ingredient that we so far neglected 
is the gauge field along time direction. Using same procedure as above, we
may define  ${\cal D}_0=\partial_t- i{\cal A}_0$ the time-like covariant 
derivative on the dual side, which shows up in commutators with $D_0$,
\bea
[D_0,Y] &\rightarrow & \Bigl({\cal F}_{01}(\sigma)\Bigr)\, 
\delta(\sigma-\sigma') , \nn
{}[D_0,X^I] &\rightarrow& \Bigl([{\cal D}_0,{\cal X}^I(\sigma)]\Bigr) \,
 \delta(\sigma-\sigma') ,\nn
{}[D_0,Z]&\rightarrow &
\Bigl({\cal D}_0(\sigma)W(\sigma,\sigma-b)\,{\cal Z} 
(\sigma-b) \nn
&&-W(\sigma,\sigma-b)\,{\cal Z} 
(\sigma-b){\cal D}_0(\sigma-b)\Bigr)  \,\delta(\sigma-\sigma'-b) .
\eea
Recall $b\equiv \epsilon/R$. ${\cal F}$ is the field strength of
$({\cal A}_0,{\cal A}_1\equiv{\cal A})$.
As evident in the above expression, the effect of the twisting is in 
the extra Wilson line factor that comes with ${\cal Z}$.
Commutators with $Y$ becomes covariant derivative along the dual circle,
\bea
[Y,X^I]  &\rightarrow& \left(i\partial_{\sigma} 
{\cal X}^I(\sigma)+[{\cal A}(\sigma),{\cal X}^I(\sigma)]\right)
\delta(\sigma-\sigma'),\nn  
{}[Y,Z]&\rightarrow&
\Bigl( [i\partial_{\sigma}+{\cal A}(\sigma)] W(\sigma,\sigma-b) 
{\cal Z}(\sigma-b)\nn
&&-W(\sigma,\sigma-b) {\cal Z}(\sigma-b){\cal A}(\sigma-b) \Bigr)\,
\delta(\sigma-\sigma'-b).
\eea
Finally commutators among ``scalar fields'' are
\bea
{}[X^I , X^K]&\rightarrow& \Bigl(
[{\cal X}^I(\sigma),{\cal X}^K(\sigma)]\Bigr) \,
\delta(\sigma-\sigma'),\nn
{}[X^I,Z]&\rightarrow&
\Bigl({\cal X}^I(\sigma)W(\sigma,\sigma-b){\cal Z}(\sigma-b)\nn
&&-
W(\sigma,\sigma-b){\cal Z}(\sigma-b){\cal X}^I(\sigma-b)\Bigr)\,
\delta(\sigma-\sigma'-b),\nn
{}[X^I,\bar Z]&\rightarrow&\Bigl({\cal X}^I(\sigma){\cal Z}(\sigma)^\dagger 
W(\sigma,\sigma+b)\nn
&&-{\cal Z}(\sigma)^\dagger W(\sigma,\sigma+b){\cal X}^I(\sigma+b)\Bigr)
\, \delta(\sigma-\sigma'+b),\nn
{}[\bar Z,Z]&\rightarrow&
\Bigl({\cal Z}^\dagger(\sigma){\cal Z}(\sigma)\nn
&&-W(\sigma, \sigma-b){\cal Z}(\sigma-b)
{\cal Z}^\dagger(\sigma-b)W(\sigma-b, \sigma)
\Bigr)\, \delta(\sigma-\sigma'),
\eea
respectively.

The action of the D0 probe descends from that of the D0 Yang-Mills
quantum mechanics whose Lagrangian is of the form,
\bea
&&\frac12\, {\rm Tr}\biggl(-\sum_A[D_0,X^A][D_0,X^A]+\sum_{A>B}
[X^A,X^B][X^A,X^B]\nn
&& \hskip 1cm +i\bar\Phi\Gamma^0[{D}_0,\Phi]  -\sum_A\bar\Phi\Gamma_A 
[{X}^A,\Phi]\biggr),
\eea
where $X^A$ includes $Y$, $Z$, $\bar Z$, as well as $X^I$.

Summations over the indices transform into 
integral over the dual circle. In the action, all commutators
must be squared and traced, but the latter operator produces a factor
of $\delta(0)$. The infinity associated with this indicates to us
that we are over-counting by treating as if all images of the D0 under
the orbifold action are real. One must divide by the total number of
D0 images, which is simply taken care of by dropping $\delta(0)$.
The Lagrangian is then integral of square of the densities appearing
the commutators above, weighted according to the D0 action. The
resulting bosonic part of the action is that of a standard $1+1$ 
dimensional Yang-Mills theories with adjoint scalars, except that
one holomorphic adjoint scalar comes with a Wilson line of length $b$ 
to the left. 

While we did not explicitly discuss fermionic quantities, they can be 
treated similarly. The only difference is that fermions rotate under 
the shift $2\pi\epsilon$ differently. Fermions will rotate under the
twist as
\be
e^{i\pi\epsilon\Gamma_{78}}
\ee
which translates to a Wilson line
\be
[P e^{-i\int_{\sigma}^{\sigma\mp b/2}{\cal A}}\,]
\ee
attached to the left of $\Psi(\sigma\mp b/2)_\pm$, where $\pm$ denotes
chirality under the $\Gamma_{78}$. Shifting the argument of quantities that
multiply $\Psi_\pm$ from the right in the same manner as above, one finds
the probe action summarized in the introduction.

\subsection{Probe Geometry}

Since Hamiltonian of the system is proportional to the
sum of absolute squares of 
the above commutators, the vacuum configuration is found by setting them
to zero. For simplicity, let us take the case of a single D0 probe and
take the temporal
gauge ${\cal A}_0=0$. Commutators with $D_0$ reduces to partial
derivatives with respect to time.

First of all, vanishing of $[Y,X^I]$ forces ${\cal X}^I$ uniform. On the
other hand, the only gauge invariant information in ${\cal A}$ at any given 
time is the Wilson line, so we may render ${\cal A}_1={\cal A}$ to be uniform 
and take value in the range of $[0, 2\pi R)$. Then, vanishing
of $[X^I,Z]$ with uniform ${\cal X}^I$, in turn forces ${\cal Z}$ to be 
uniform also. Thus, we seem to find vacuum moduli space of the form,
\be
{\bf R}^6\times {\bf R}^2 \times{\bf S}^1 .
\ee
However, this is only a topological statement and the actual moduli space
metric is determined by slow motion along the moduli space. Recalling that
we adopted the temporal gauge ${\cal A}_0=0$, we may 
write these moduli as
\bea
{\cal X}^I &= &  x^I ,\nn
{\cal Z}&=&  r e^{i \theta}, \nn
{\cal A}&=&  y ,
\eea 
from which it also follows that ${\cal F}_{01}=\partial_0 y$. The only
nontrivial aspect of the kinetic terms for these fields is that the Wilson 
line mixes with the phase of ${\cal Z}$, and in fact the two
always appears in the combination of $ \theta + b\,  y$. In effect, 
the moduli space metric, which is extracted from time derivative part 
of the kinetic terms, is
\be
(d x^1)^2+\cdots+(d x^6)^2+
d r^2+ r^2(d \theta +b\, d y)^2
 +(d y)^2 ,
\ee
reproducing the spacetime that is being probed by the D0 brane.

\subsection{A Consistency Check: T-duality}

On the other hand, T-duality takes the flat 
Melvin background to one with  a nontrivial spacetime metric, 
dilaton, and antisymmetric tensor  \cite{russo2},
\be
g+B= -(dt)^2+(dx^1)^2+\cdots+(dx^6)^2+dr^2+\frac{r^2}{\Lambda}
(d\theta- b\, d\tilde y)(d\theta+ b\, d\tilde y) +(d\tilde y)^2 ,
\ee
where $\tilde y$ parameterizes the dual circle. Curved nature of this
background is summarized in the function, 
\be
\Lambda\equiv 1+b^2r^2 ,
\ee
which also enters the dilaton:
\be
e^{2(\Phi_0-\Phi)}=\Lambda .
\ee
Since we could consider the D0 probe above  as a 
D-string probing the dual background, one might wonder why the nontrivial 
curvature of the dual geometry is not seen by the probe. Here, we will
clarify this issue by considering D-string effective action in the dual
background.

A single  D-string probing this geometry is governed by a Born-Infeld action,
\be
e^{-\Phi}\sqrt{-{\rm Det}(g+B+\cal F)} .
\ee
T-dual of the D0 would be a D-string wrapped around $\tilde y$. Set the
worldvolume coordinate such that
\be
t(\tau,\sigma)=\tau,\qquad \tilde y(\tau,\sigma)=\sigma ,
\ee
and estimate the determinant up to two derivative terms. 
\bea
&&-\frac{1}{\Lambda}\nn
&&+ \frac{1}{\Lambda}\left((\partial_0 \vec x)^2
+(\partial_0 r)^2 +({r^2}/{\Lambda})(\partial_0 \theta)^2\right)
-\left((\partial_1 \vec x)^2+(\partial_1 r)^2
+({r^2}/{\Lambda})(\partial_1 \theta)^2\right)\nn
&&+
\left({\cal F}_{01}+b{r^2} \partial_0\theta/{\Lambda}\right)^2+\cdots .
\eea
Taking into account the dilaton factor, we find the following leading
kinetic terms from derivative expansion of Born-Infeld action,
\bea
&&-\frac{1}{2}({\cal F}_{01})^2 -\frac{1}{2}\left((\partial_0 \vec x)^2
+(\partial_0 r)^2 +r^2(\partial_0 \theta+b{\cal F}_{01})^2\right) \nn
&&+\frac{1}{2}\left((\partial_1 \vec x)^2+(\partial_1 r)^2
+{r^2}(\partial_1 \theta)^2\right)
+\frac{1}{2}b^2r^2\left((\partial_1 \vec x)^2+
(\partial_1 r)^2\right) ,
\eea
up to an overall factor of $e^{-\Phi_0}$. 

Now it is a matter of
straightforward exercise to check the same action results from the
$U(1)$ probe theory. For instance, the first line reproduces
the probe metric. The mixing between $\partial_0\theta$
and ${\cal F}_{01}$ is precisely what happened from the gauge theory side.
The last two terms may seem unfamiliar. But, recalling that we performed
a derivative expansion on Born-Infeld side, we should also take the limit
where $\partial_1$ derivatives of fields $\cal X$ and $\cal Z$ are small. 
In this limit, two sets of nontrivial commutator terms are
\bea
\frac12 {\rm Tr}[X^I,Z][X^I,\bar Z] &\rightarrow& \frac12\int d\sigma \,
{\cal Z}(\sigma)\bar{\cal Z}(\sigma)
({\cal X}^I(\sigma+b)-{\cal X}^I(\sigma))^2\nn
&&\simeq \frac12\int d\sigma \,
{\cal Z}(\sigma)\bar{\cal Z}(\sigma)\left(b^2
(\partial_1 {\cal X}^I(\sigma))^2\right)\nn
 &\rightarrow& \frac12\int d\sigma \,b^2r^2(\partial_1 \vec x)^2 ,\nn
&&\nn
\frac12 {\rm Tr}([Z,\bar Z]/2) ([\bar Z, Z]/2)
&\rightarrow& \frac12\int d\sigma \,
\left([{\cal Z}^\dagger(\sigma){\cal Z}(\sigma)-{\cal Z}(\sigma-b)
{\cal Z}^\dagger(\sigma-b)]/2\right)^2 \nn
&&\simeq \frac12\int d\sigma \,\left(b\,
\partial_1({\cal Z}\bar{\cal Z})/2\right)^2 \nn
&\rightarrow& \frac12\int d\sigma \,b^2r^2(\partial_1 r)^2 ,
\eea
which results from the nonlocal nature of each term.
The upshot is that while the full Born-Infeld action knows about the
nontrivial geometry, one must truncate to two-derivative terms in
order to make comparison with the gauge theory result. At this
level, the two approaches produce identical results.

\section{${\bf C}/Z_n$ Orbifold Limit and Closed String Tachyon}

It has been noted that, when $\epsilon=1+1/n$ with odd integer $n$, 
the above orbifold  in the limit $R\rightarrow 0$, reduces to a singular 
orbifold ${\bf C}/Z_n\times {\bf R}^1$ \cite{suyama,takayanagi1}. \footnote{
This is not obvious from 
the background geometry itself, but was rather deduced from the 
form of perturbative string spectrum. We suspect that some nontrivial
field redefinition is at work, in much the same way as in
Ref.~\cite{gutperle1}.} In this section, this reduction
to the orbifold limit is also confirmed from the D-brane probe side by
making contact with the probe theory found by Adams et.al. This 
comparison should, in part, serve as a useful tool for understanding 
how tachyon couples to the probe dynamics.

\subsection{Orbifold Limit of Probe Theory}

The probe theory of the ${\bf C}/Z_n$ orbifold is a $U(N)^n$ quiver theory
\cite{adams}.
One may wonder how the present $U(N)$ theory can ``reduce'' to a theory
with apparently larger gauge group. The keypoint is that we may think of
the case $\epsilon=1+1/n$ as
\be
({\bf R}^2\times {\bf S}^1)/Z_n
\ee
where ${\bf S}^1$ is a circle of $nR$. The dual circle is then of 
circumference $1/nR$, which may be thought of as a $1/n$ part of 
the dual circle we used above.
In the following, each factor of $U(N)^n$ will be simply $U(N)$ restricted
to each of these intervals of length $1/nR$. 
Cutting the dual circle of length $1/R$
into $n$ intervals of length $b=1/nR$ each, say,
$[-1/2nR,1/2nR)$, $[1/2nR,3/2nR)$, $[3/2nR,5/2nR)$, etc,   each of these 
intervals grows to be an infinite line of its own. 

Instead of working with $n$ copies of a real line, a more sensible thing to do 
is treat the dynamical fields in each segment line as a field of its own.
Namely split each of local field, say $\cal A$,  into $n$ independent 
quantities ${\cal A}_a$ with a cyclic label $a=1,\dots, n, n+1\sim 1$, 
such that
\be
{\cal A}_a(\sigma)=\lim_{R\rightarrow0}{\cal A}(\sigma+a/nR) .
\ee
Here we invoked the periodicity of the dual circle, $1/R$, i.e.,
$\sigma\sim \sigma+1/R$ on the right hand side prior to taking the limit.
The range of $\sigma $ in this definition is  restricted to be
between $-1/2nR$ and $1/2nR$, 
which eventually becomes the entire real line upon the limit.
This gives us $n$ gauge fields that propagate along ${\bf R}^1$ of 
${\bf C}/Z_n\times {\bf R}^1$.
The net effect is to have $U(N)_1\times U(N)_2 \times  \cdots \times  U(N)_n $ 
gauge group instead of $U(N)$. 
Similar decomposition works for ${\cal X}^I$, 
\be
({\cal X}^I)_a'(\sigma)=\lim_{R\rightarrow0}{\cal X}^I(\sigma+a/nR) .
\ee
giving us six adjoint fields for each $U(N)$ factor. 

For ${\cal Z}_a$'s, we want to include the Wilson line intact, so that
\be
{\cal Z}_a'(\sigma)=\lim_{R\rightarrow0}{W}(\sigma+a/nR,\sigma+(a-1)/nR)
{\cal Z}(\sigma+(a-1)/nR),
\ee
transforms as a bifundamental under $U(N)_a\times  U(N)_{a-1}$ and is
effectively a local field in this limit. 
Finally fermions come with a Wilson line of length $\mp \epsilon/2R=
\mp (n+1)/2nR$.
For example,
\bea
(\Psi_\pm)_a'(\sigma)=\lim_{R\rightarrow0}{W}(\sigma+a/nR,
\sigma+(a\mp(n+1)/2)/nR)&&\nn
\times \Psi_\pm(\sigma+(a\mp(n+1)/2)/nR  )&& \hskip -5mm,
\eea
are each in the bifundamental representation under $U(N)_a\times 
U(N)_{a\mp(n+1)/2}$. Note that $(n+1)/2$ is always an integer 
since $n$ is an odd integer. The combined field content is exactly that of 
the probe theory on ${\bf C}/Z_n$ as found by Adams et.al. \cite{adams}.

\subsection{Reconstructing Twisted Sector Tachyons}

Coming back to the case of $N=1$, 
an interesting observation Adams et.al. made is that closed string tachyons
couple to these worldvolume fields in a manner reminiscent of 
Fayet-Illiopoulos terms in probe theories for supersymmetric
orbifolds. For $U(1)^n$ case, the leading coupling can be rewritten as
\be
\sum_a \zeta_a(|{\cal Z}_a'|^2-|{\cal Z}_{a+1}'|^2), \label{FI}
\ee
where parameters $\zeta_a$ obey $\sum_a\zeta_a=0$ and 
are turned on by the condensation of tachyons  \cite{adams,harvey}.
The role of such $\zeta$'s was seen to resolve the singularity
at the conical tip of ${\bf C}/Z_n$, which leads them to conjecture that
the condensation of tachyons leads the theory back to some supersymmetric
critical string theory.

The closed string tachyons of the ${\bf C}/Z_n$
 orbifold reside in $n-1$ twisted sectors
of the orbifold. The vacuum energy of the $k$-th twisted sector 
(in Green-Schwarz formalism) is
\bea
-\frac{k}{2n} & & \hbox{for $k$ even },\nn
-\frac{n-k}{2n} & & \hbox{for $k$ odd },
\eea
which, given an odd $n$, produces a pair of tachyons for each value of 
mass squared 
\be
-\frac{l}{n}\times\frac{4}{\alpha'}, \qquad l=1,2,\dots, [n/2] .
\ee
where we restored the explicit factor of $\alpha'$.
These tachyons live in ${\bf R}^{1+7}$ and, in particular, propagate
along the $\sigma$ direction arising from the dual circle.

{}From the Melvin viewpoint these tachyons can be reconstructed
as follows. In the limit of $R\rightarrow 0$, a sector with 
the winding number $w$ has the vacuum energy that depends on
non-integer part of $\epsilon w$ (in NSR formalism),\footnote{
This vacuum energy is with respect to the Fock vacuum 
defined by $\psi_r|0\rangle
=\psi^\dagger_{-r}|0\rangle=\widetilde{\psi}_r|0\rangle=
\widetilde{\psi}^\dagger_{-r}|0\rangle=0\ (r>0)$, where $\psi_r$ and 
$\widetilde{\psi}_r$ are modes of the free worldsheet fermions with 
the twisted boundary condition\cite{russo2}:
$\psi(z)=\sum_r\psi_r z^{-r-1/2},
\widetilde{\psi}(\bar{z})=\sum_r\widetilde{\psi}_r\bar{z}^{-r-1/2}$
}
\be
-\frac 12(\epsilon w-[\epsilon w]) ,
\ee
or
\be
-\frac 12+\frac 12(\epsilon w-[\epsilon w]) ,
\ee
depending on whether the quantity inside the parenthesis is smaller or 
larger than $1/2$. One must further take into account the GSO projection
which changes depending on whether the integer part of $\epsilon w$ is odd 
or even.\footnote{
The mass squared of the lowest mode with winding number $w$ with finite $R$ 
is 
 \cite{russo2,suyama}
\be
\left\{\begin{array}{ll}
(\frac{wR}{\alpha'})^2-\frac{2}{\alpha'}(\epsilon w-[\epsilon w]) &
\mbox{for } [\epsilon w]=\mbox{even}, \\
(\frac{wR}{\alpha'})^2-\frac{2}{\alpha'}(1-(\epsilon w-[\epsilon w])) &
\mbox{for } [\epsilon w]=\mbox{odd},
\end{array}\right.
\ee
where the first term is the contribution of winding number, and the 
second term is the sum of zero point energy and oscillator number which
gives the result of (\ref{tachyonmass}).
Tachyonic region for $w\not\neq 1$ modes is contained by that of $w=1$.
If we take $\epsilon=1+\frac{1}{n}$, $w=1$ mode becomes massless when 
$R=\sqrt{2\alpha'(1-1/n)}$.}
The lowest lying state after GSO projection gets contribution 
from left(right) moving oscillators as follows. With $\epsilon =1+1/n$, 
\be
-\frac{2l}{2n}
\ee
for $w=2l$ and
\be
-1+\frac{(n-2l)}{2n}=-\frac{2l}{2n}
\ee
for $w=n-2l$. The pattern repeats itself for $w=kn\pm 2l$ for all
integer $k$. Thus there are two tachyonic towers for each integer 
$l$ less than $[n/2]$, whose mass squared is
\be
-\frac{l}{n}\times\frac{4}{\alpha'}, \qquad l=1,2,\dots, [n/2] .
\label{tachyonmass}
\ee
On the T-dual side, the winding numbers translate to quantized momenta 
along the dual circle. Successive momentum modes in any given tower are
separated by $n$ basic quanta, so each tower is appropriate for building up a 
complete field along a $1/n$ segment of the dual circle. Thus, we find
$n-1$ tachyonic fields that may propagate along ${\bf R}^1$ of 
${\bf C}/Z_n\times {\bf R}^1$.
 This agrees with findings by Adams et.al. 
Also, it is fairly obvious that the winding number $w$  modulo $n$
labels the $n-1$ twisted sectors in this orbifold limit.

\section{Conclusion}

We have considered D0 brane probe in the background associated with 
Melvin type twisting. The probe dynamics is written down 
explicitly, and is almost identical to a maximally supersymmetric 
$U(N)$ theory except that a complex adjoint Higgs and all fermions 
are accompanied by a open Wilson line. This renders the Lagrangian 
density to be a nonlocal expression. The flat directions are studied 
and the probe geometry has been shown to agree with the background, 
where the open Wilson line plays a crucial role. We have further 
considered a $Z_n$ orbifold limit that has been argued to 
reduce to ones in Ref. \cite{adams}, and reproduced the probe 
dynamics of the latter by taking an appropriate limit. 

An important question that was not addressed in this note is how 
the closed string tachyon vev couples to the probe 
in general Melvin background. Among the constraints the coupling must
satisfy is that it should reproduce that of Adams et.al. in the 
orbifold limit. For instance, if a term of type
\be
\int d\sigma\; \zeta(\sigma)\times\left(|{\cal Z}(\sigma)|^2
-|{\cal Z}(\sigma-b)|^2\right) ,
\ee
with constraint 
\be
\int d\sigma \zeta(\sigma) =0.
\ee
is present for a single D0 probe, it will reproduce (\ref{FI}) in the 
${\bf C}/Z_n$ limit. 
In fact, if $\zeta$ is linear in the tachyon, this constraint 
is automatically satisfied since tachyon vev must have nontrivial momentum 
along the dual circle. For any finite $R$, at most a finite number of winding
modes will be tachyonic and the profile of condensed tachyon is never
uniform along the dual circle. What must happen in the 
${\bf C}/Z_n$ orbifold limit is that an infinite number of winding modes 
become tachyonic, so a piece-wise uniform tachyon vev 
becomes possible along the dual circle and produces $\{\zeta_a\}$. 
Computation of the precise coupling of 
tachyon is beyond the scope of this note, however, and we would like to 
come back to the issue of tachyon condensation elsewhere \cite{later}.

\vskip 1cm
We would like to thank T. Takayanagi for a useful discussion. At the 
conclusion of this work, several papers appeared that address related 
unstable theories \cite{vafa,harvey,dabholkar,shiraz}

\end{document}